\documentclass{iau} 
\usepackage{graphicx}

\title[Molecular complexity in the ISM] 
{Molecular complexity\\ in the interstellar medium}

\author[Arnaud Belloche]   
{Arnaud Belloche$^1$
}

\affiliation{$^1$Max-Planck-Institut f\"ur Radioastronomie,\\ Auf dem H\"ugel 69, 53121 Bonn, Germany \\ email: {\tt belloche@mpifr-bonn.mpg.de} \\[\affilskip]
}

\pubyear{2019}
\volume{xxx}  
\jname{Laboratory Astrophysics: from Observations to Interpretation}
\editors{XXX, eds.}
\begin{document}

\maketitle

\begin{abstract}
The search for complex organic molecules in the interstellar medium (ISM) has 
revealed species of ever greater complexity. This search relies on the 
progress made in the laboratory to characterize their rotational spectra. Our 
understanding of the processes that lead to molecular complexity in the ISM 
builds on numerical simulations that use chemical networks fed by laboratory 
and theoretical studies. The advent of ALMA and NOEMA has opened a new door 
to explore molecular complexity in the ISM. Their high angular resolution 
reduces the spectral confusion of star-forming cores and their increased 
sensitivity allows the detection of low-abundance molecules that could not be 
probed before. The complexity of the recently-detected molecules manifests 
itself not only in terms of number of atoms but also in their molecular 
structure. We discuss these developments and report on ReMoCA, a new spectral 
line survey performed with ALMA toward the high-mass star-forming region 
Sgr B2(N).


\end{abstract}

\firstsection 
\section{Growing complexity of interstellar molecules}

A complex organic molecule (COM), that is, an organic molecule with six atoms 
or more according to the definition currently adopted in astrochemistry 
(\cite[Herbst \& van Dishoeck 2009]{Herbst09}), was among the first ten 
molecules identified in the ISM: methanol, CH$_3$OH, was detected
the same year as carbon monoxide, CO (\cite[Ball et al. 1970]{Ball70}, 
\cite[Wilson et al. 1970]{Wilson70}). Since then, with the continuous 
developments of radio astronomy and the progress made in the 
laboratory to characterize the rotational spectra of molecules, 
74 COMs have been identified in the ISM as of June 2019 (see, e.g., 
``Molecules in Space'' in the Cologne Database for Molecular 
Spectroscopy\footnote{https://cdms.astro.uni-koeln.de/classic/molecules},
and \cite[McGuire 2018]{McGuire18b}). They represent about one third of the
total number of known interstellar molecules (212). Excluding the 
fullerenes, the largest molecule identified in the ISM so far, 
\textit{c}-C$_6$H$_5$CN 
(see Sect.~\ref{ss:aromatic}), has 13 atoms. However, much larger molecules 
have been identified in meteorites, with for instance more than 80 different 
amino acids (see, e.g., \cite[Botta \& Bada 2002]{Botta02}). The recent 
in-situ chemical inventory of comet 67P/Churyumov-Gerasimenko by the 
\textit{Rosetta} mission has also revealed molecules such as glycine that have 
not been detected in the ISM yet (\cite[Altwegg et al. 2016]{Altwegg16}, 
\cite[2017]{Altwegg17}). This gives us good 
reasons to believe that the number of molecules present in the ISM is much 
larger than what we know today.

One of the goals of astrochemistry is to explore the degree of chemical  
complexity that can be reached in the ISM. Beyond the mere identification of
molecules, astrochemists want to understand the processes that lead 
to the formation of COMs in the ISM. It is also important to
investigate how chemical complexity is passed over (or not) from one stage of
star formation to the next one. This will tell us if the molecular 
complexity of meteorites and comets in our solar system is a widespread 
outcome of interstellar chemistry in our Galaxy. To make progress, 
a close interplay between observations, astrochemical
modelling, and experiments is necessary. We discuss below a few recent results 
that illustrate the growing complexity of interstellar molecules in terms of 
molecular structure. Sections~\ref{s:ReMoCA} and \ref{s:urea} present a new 
ALMA spectral survey and one of its first results.

\subsection{Chiral molecules}
\label{ss:chiral}

Life on Earth is based on homochirality, with amino acids being left-handed and
sugars being right-handed. The origin of this homochirality might be 
extraterrestrial (\cite[Bonner et al. 1999]{Bonner99}). Several unsuccessful 
searches for chiral molecules in the ISM were reported in the past: 
\cite{Jones07} and \cite{Cunningham07} did not detect propylene oxide, 
\textit{c}-CH(CH$_3$)CH$_2$O,
toward Sgr~B2(N) and Orion-KL with the Australia Telescope Compact Array and 
the Mopra telescope, respectively; \cite{Mollendal12} failed to detect 
2-aminopropionitrile, CH$_3$CH(NH$_2$)CN, toward Sgr~B2(N) with the IRAM 30\,m 
telescope, and \cite{Richard18} did not succeed with ALMA either. Some progress
was made with ALMA by \cite{Belloche16} with a tentative detection toward 
Sgr~B2(N) of deuterated ethyl cyanide, CH$_3$CHDCN, which is chiral through the deuterium substitution. The first secure detection of a 
chiral molecule in the ISM came shortly after, with three transitions of 
propylene oxide detected in absorption toward Sgr B2(N) by \cite{McGuire16} 
with the Green Bank Telescope (GBT) and the Parkes telescope. Although the
number of detected lines is small for the identification of such a complex 
molecule, the low spectral confusion in the cm range gives
confidence in the assignment. This radio detection does not tell if 
propylene oxide has an enantiomeric excess, but \cite{McGuire16} 
discuss in detail the prospects of such a measurement using circular 
dichroism. The recent identification of two chiral derivatives of propylene 
oxide by \cite{Pizzarello18} in the Murchison meteorite, both with an 
enantiomeric excess, suggests that an enantiomeric excess in interstellar 
propylene oxide is plausible.

\subsection{Aromatic molecules}
\label{ss:aromatic}

Polycyclic aromatic hydrocarbons (PAHs) are thought to be the carriers of the
unidentified infrared bands (\cite[Tielens 2008]{Tielens08}), but none
has been firmly identified so far. After the
uncertain detection of benzene by \cite{Cernicharo01}, which relied on only 
one weak mid-infrared absorption feature in its $v_4$ bending mode, the recent 
GBT detection of the aromatic molecule benzonitrile, \textit{c}-C$_6$H$_5$CN, 
toward the cold dense core TMC-1 by \cite{McGuire18a} may represent a key step 
to understand the formation of polycyclic aromatic molecules in the ISM. Eight 
rotational transitions were detected in emission, some of them even with their 
hyperfine structure resolved, yielding a robust identification. 
\cite{McGuire18a} proposed two possible formation routes for benzene and/or 
benzonitrile, one involving electron irradiation of acetylene, in the gas 
phase or in the icy mantles of dust grains (\cite[Field 1964]{Field64}, 
\cite[Zhou et al. 2010]{Zhou10}), and the other one via a cyclization process
of large cyanopolyynes HC$_n$N (\cite[Loomis et al. 2016]{Loomis16}). 
Benzonitrile may also form from the addition of CN to benzene, which makes it 
an interesting tracer of benzene, a molecule from which PAHs are expected to 
form (\cite[Joblin \& Cernicharo 2018]{Joblin18}).

\subsection{Branched molecules}
\label{ss:branched}

Our earlier spectral survey of Sgr~B2(N) with the IRAM 30\,m telescope
(\cite[Belloche et al. 2013]{Belloche13}) led to the detection of several new 
COMs, in particular \textit{normal}-propyl cyanide, \textit{n}-C$_3$H$_7$CN 
(\cite[Belloche et al. 2009]{Belloche09}). With the advent of ALMA, we started
a project to explore the molecular complexity of Sgr~B2(N) at higher angular
resolution. We made significant progress with the first survey called EMoCA 
(Exploring Molecular Complexity with ALMA) at an angular resolution of 1.5'',
which allowed us to obtain individual spectra of the main hot cores,
N1 and N2, that were blended in the single-dish beam.
Sgr~B2(N2) was found to have narrow linewidths (5~km~s$^{-1}$), which 
means a reduced spectral confusion. Among other results, EMoCA led to the
first detection of a branched alkyl molecule, \textit{iso}-propyl cyanide, 
\textit{i}-C$_3$H$_7$CN, a structural isomer of \textit{n}-C$_3$H$_7$CN 
(\cite[Belloche et al. 2014]{Belloche14}). The detection of a branched molecule
opens a new domain in the structures available to the chemistry of star-forming
regions. It also establishes a further link between interstellar chemistry and
the chemical composition of meteorites, where branched molecules even dominate
over straight chain ones (\cite[Cronin \& Pizzarello 1983]{Cronin83}).

\section{The ReMoCA survey}
\label{s:ReMoCA}

The chemical models presented in \cite{Belloche14}, using the astrochemical 
code MAGICKAL (\cite[Garrod 2013]{Garrod13}), reproduce the abundance ratio 
\textit{i}/\textit{n} of 0.4 measured for propyl cyanide, and further 
simulations by \cite{Garrod17} predict 
that for butyl cyanide, C$_4$H$_9$CN, one of the branched isomers should even 
dominate over the straight chain form. These predictions and the rotational
spectroscopy work performed on these isomers in parallel 
(\cite[M\"uller et al. 2017]{Mueller17}, \cite[Wehres et al. 2018]{Wehres18})
motivated us to search for the isomers of butyl cyanide and perform the ReMoCA 
survey (Re-exploring Molecular Complexity
with ALMA) toward Sgr~B2(N) during ALMA's Cycle 4, improving by a factor of 
three both the angular resolution and the sensitivity compared to EMoCA. The 
high angular resolution of 0.5'' was key to resolve the emission 
around the main hot core Sgr~B2(N1). While the peak position of Sgr~B2(N1) is 
affected by the high optical depth of the dust emission, the high angular 
resolution allowed us to spot slightly offset positions with narrow linewidths 
(5~km~s$^{-1}$). We thus now have the possibility to characterize the chemical 
composition of Sgr~B2(N1) at unprecedented sensitivity.

\section{Detection of urea}
\label{s:urea}

One of the first results of the ReMoCA survey is the detection of urea,
NH$_2$C(O)NH$_2$, toward Sgr~B2(N1) 
(\cite[Belloche et al. 2019]{Belloche19}). Urea
was discovered in the 18$^{\rm th}$ century as an \textit{organic} (that is, in 
the paradigm of that epoch, produced by living organisms) molecule. It was
synthetized by \cite{Woehler28} from \textit{inorganic} compounds, which was
a key experiment now regarded as the start of modern organic chemistry.
Urea was detected in meteorites by \cite{Hayatsu75} but two inconclusive 
searches for it in the ISM were reported 
(\cite[Raunier et al. 2004]{Raunier04}, 
\cite[Remijan et al. 2014]{Remijan14}). The ReMoCA identification of urea 
relies on nine lines clearly detected toward Sgr B2(N1), in its ground and
first vibrational states. The recent rotational spectroscopy characterization 
of its vibrational states by \cite{Thomas14} and \cite{Kisiel14} was essential 
to secure the detection. We find that urea is one and two orders of magnitude 
less abundant than acetamide, CH$_3$C(O)NH$_2$, and formamide, NH$_2$CHO, 
respectively. Surprisingly, it is at least one order of magnitude less 
abundant in Sgr~B2(N2) than in Sgr~B2(N1), relative to formamide. The reasons 
for this difference between the two sources will have to be explored.

\section{Conclusions}
\label{s:conclusions}

The recent advances in the search for COMs in the ISM have revealed that 
interstellar chemistry is capable of producing chiral, aromatic, and branched 
molecules. These molecules are still relatively simple (less than 13 atoms), 
but their molecular structures are reminiscent of the structures that 
characterize more complex molecules in meteorites. Astronomers have
thus probably probed only a small fraction of the molecules
present in the ISM. The large partition functions of more complex species
may ultimately hamper their detection, but the molecules detected
so far, with their variety of molecular structures, already allow us to 
identify with numerical simulations which chemical processes are key to 
produce these structures. On the observational side, ALMA and NOEMA have only
started to deliver the chemical composition of star-forming regions and we
can expect further progress in the coming decade. On a longer term,
ngVLA and SKA may also contribute to push the limits of chemical complexity in 
the ISM even further.


\end{document}